# Benchmarking the performance of DFT functionals for absorption and fluorescence spectra of EGFR inhibitor AG-1478 using TD-DFT

*Sallam Alagawani[1], Vladislav Vasilyev[2] and Feng Wang[1]\**

[1]Department of Chemistry and Biotechnology, School of Science, Computing and Engineering Technologies, Swinburne University of Technology, Melbourne, Victoria 3122, Australia

[2]National Computational Infrastructure, Australian National University, Canberra, ACT 0200, Australia

*fwang@swin.edu.au

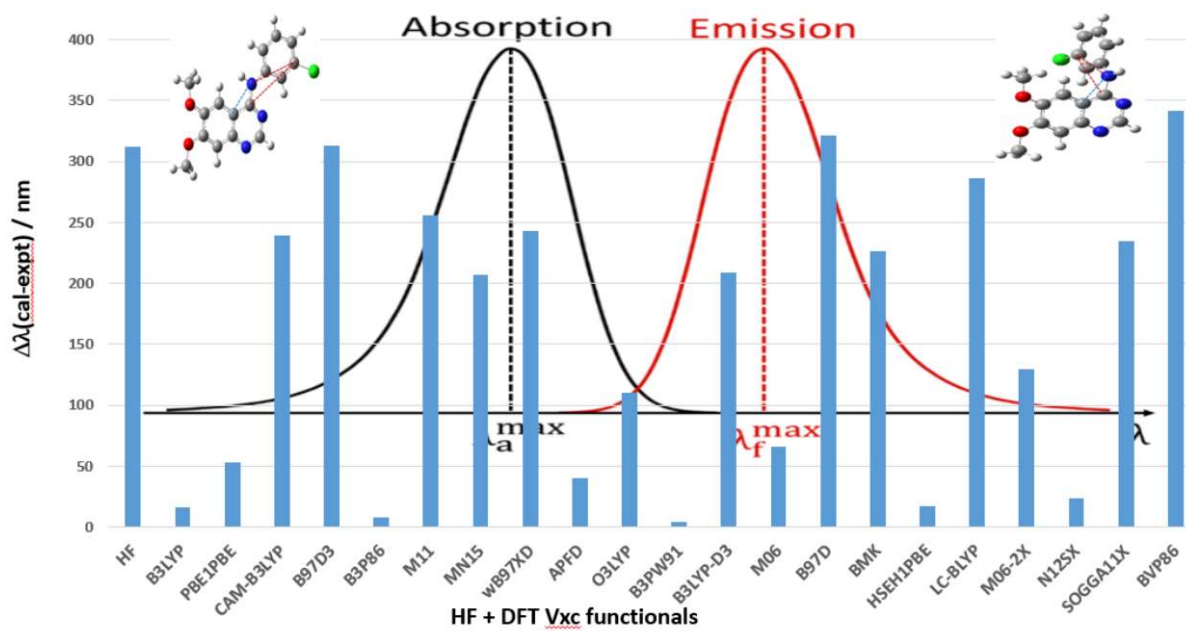

**ABSTRACT**

Optical spectra (UV-vis and fluorescence) are sensitive to the chemical environment and conformation of fluorophores and therefore, are ideal probes for their conformation and solvent responses. Tyrosine kinase inhibitors (TKI) such as AG-1478 of epidermal growth factor receptor (EGFR) when containing a quinazolinamine scaffold are fluorophores. Conformers of AG-1478 have been confirmed by both experiment and theory. It is, however, very important to benchmark computational methods such as DFT functionals against optical spectral measurements, when time-dependent density functional theory (TD-DFT) is applied to study the fluorophores. In this study, the performance of up to 22 DFT functionals is benchmarked with respect to the measured optical spectra of AG-1478 in dimethyl sulfoxide (DMSO) solvent. It is discovered when combined with the 6-311++G(d,p) basis set, B3PW91, B3LYP, B3P86, BPE1BPE, APFD, HSEH1PBE, and N12SX DFT-$V_{XC}$ functionals are the top performers. Becke's three-parameter exchange functional (B3) tends to generate accurate optical spectra over other exchange functionals. The B3PW91 functional is the recommended DFT functional for optical property calculations of this class of TKIs whereas B3LYP is excellent for absorption calculations and B3P86 is best for emission calculations. Any corrections to B3LYP, such as CAM-B3LYP, LC-B3LYP, and B3LYP-D3 result in larger errors in the optical spectra of AG-1478 in DMSO solvent. These B3Vc functionals are reliable tools for optical properties of the TKIs and therefore the design of new agents with larger Stokes shift for medical image applications.

**1. INTRODUCTION**

The epidermal growth factor receptor (EGFR) as a membrane-spanning cell surface protein is a major target for drugs in treating lung carcinoma.[1] Mutation in the tyrosine kinase domain of EGFR often arises in human cancers, which can cause drug sensitivity or resistance by influencing the relative strengths of drug and adenosine triphosphate (ATP) binding. To understand the mechanism of such mutations dysregulate the EGFR and to modulate its sensitivity and resistance to tyrosine kinase inhibitors (TKIs), the most powerful solution is to combine experimental and theoretical means at the molecule level.[2] The scaffold with the 4-quinazolinamine class of TKIs is excellent fluorophores for optical probes of the interaction of anticancer drugs and the environment when interacting with light. Optical spectral (UV-vis and fluorescence) properties of quinazolinamine fluorophores are sensitive to the chemical environment and conformation of the fluorophores and therefore, are ideal tools for optical reporting of inhibitors such as AG-1478 of protein tyrosine kinases EGFR.

As a highly potent TKI, N-(3-chlorophenyl)-6,7-dimethoxy-4-quinazolinamine (AG-1478) has been studied for its conformation in solvents and for its optical spectra using combined experimental and theoretical methods.[3-5] It has been found that compared to the absorption UV-vis spectrum, the fluorescence spectrum of AG-1478 is more sensitive to the solvent environment[4] and conformer presence.[3,5] This property makes the TKIs useful as probes of electronically excited states in biochemical studies, which has opened new possibilities for elucidating molecular mechanisms of photobiological processes.[6] For example, in a recent computational study concerning the unsubstituted flavone molecule, Marian et al.[7] showed that the first excited singlet state favours a planar arrangement of the phenyl and chromone rings, in contrast to the twisted conformation adopted in the electronic ground state.

Next, Stokes shift is an important property for bioactive compounds (drugs) including fluorescent dyes. The fluorophores are used in a broad spectrum of in vitro biological applications owing to their optical and structural tenability, non-invasive treatment, cell compatibility, and real-time response.[8] Quinazoline scaffold-based TKIs are also ideal fluorescent dyes due to its large Stokes shift ($\lambda_{ss}$ at 129.5 nm of AG-1478 in DMSO solvent),[4] which is significantly larger than many of the widely adopted fluorescent dyes, such as fluorescein, rhodamine, oxazine, and cyanine with $\lambda_{ss} \sim 30$ nm.[9] Fluorescent dyes with small Stokes shift result in poor signal-to-noise ratio and self-quenching on present microscope configurations. Stokes shifts of fluorophores depend on the accuracy of both the absorption and emission, providing additional information.

Optical spectra of electronic systems can be calculated quantum mechanically using the frequency-dependent linear response (LR) time-dependent density functional theory (TD-DFT) method.[10] TD-DFT is an exact theory that relies on the analysis of the TD LR of the exact ground-state density to a TD external perturbation, which after Fourier transformation yields exact excited-state energies and oscillator strengths. The derivation of the Runge-Gross theorem and the subsequent formulation of a TD Kohn-Sham equation were the cornerstones in the development of the TD-DFT formalism, which becomes the most prominent method for the calculation of excited states of medium-sized and large molecules since its development in 1984.[11] The majority of TD-DFT applications are carried out using the vertical approximation. In this framework, one estimates the excited state (ES) energies (and related properties) using the optimal ground-state (GS) geometry, without exploring the ES potential energy surface. This approach can be a good approximation for absorption (e.g. UV-vis) spectral calculations. Though TD-DFT is an exact theory, it is based on the adiabatic approximation (frequency independence) and since the exact exchange-correlation ($V_{XC}$) functional is not known, approximate $V_{XC}$ functionals need to be employed in a practical calculation. Both limit the accuracy of TD-DFT calculations.[12-13] The functionals such as B3LYP and PBE have been the most widely used $V_{XC}$ functionals in standard ground-state DFT applications.[11] The selection of an adequate exchange-correlation functional ($V_{XC}$) for modelling excited state properties, however, has been the subject of many benchmark studies.

Although general conclusions have been made, there is not universally "best" $V_{XC}$ for TD-DFT in all drug/molecular systems. The TD-DFT benchmark studies need to be made for a specific class of molecular systems. For example, in the benchmarking study for up to six hybrid DFT functionals (B3LYP, PBE0, M06, M06-2X, CAM-B3LYP, and LC-PBE), Charaf-Eddin et al.[12] assessed the DFT functionals in light of the experimental band shapes corresponding to both the absorption and emission spectra of a set of 20 representative conjugated compounds. They found that all these tested functionals but LC-PBE reproduce the main experimental features for both absorption and fluorescence of the conjugated molecules. They further noticed that the M06-2X functional provides accurate excitation energies for problematic molecules. More recently, Bay et al.[14] benchmarked up to 11 DFT $V_{XC}$ functionals, for their performance of maximum absorption wavelengths ($\lambda_{max}$) of coumarin derivatives. They discovered that the B3LYP and APFD DFT functionals topped the list of high performance. Shao et al.[6] benchmarked up to 17 commonly used DFT $V_{XC}$ functionals for 11 green fluorescent protein (GFP) chromophore models such as p-hydroxybenzylidene imidazolinone models and the photoactive yellow protein (PYP) chromophore models such as p-vinyl phenol (pVP) and trans-p-coumaric acid (pCA). The TD-DFT calculated vertical excitation energies of the five lowest excited singlet states were compared with calculations using the approximate second-order coupled-cluster theory level (CC2) with def2-TZVP and aug-def2-TZVP basis sets.[6] When using the aug-def2-TZVP basis set, they found that B3LYP, PBE0, and M06-2X functionals yield similar results as obtained with the range-separated functionals (CAMB3LYP, CAMh-B3LYP, ωPBE, ωhPBE0, ωPBEh, and ωB97X-D) for the lowest excited singlet state (S1).[6]

In many cases, results obtained with TD-DFT are quite sensitive to the choice of the $V_{XC}$ functionals. Therefore, the reliability of TD-DFT calculations should always be checked by comparison with either wave-function-based benchmark calculations or experimental data, as well as by the sensitivity of the results to the choices of $V_{XC}$ functional.[15] In the present study, we benchmark 22 DFT $V_{XC}$ functional for the maximum absorption and emission wavelengths, as well as the Stokes shift of AG-1478 ($\lambda_{max}$) with respect to available solution experimental optical spectra of AG-1478 in solution.[3-4] The top DFT $V_{XC}$ performers will be applied to study other TKIs sharing the same scaffold.

## 2. METHODS AND COMPUTATIONAL DETAILS

All ground state quantum mechanical optimization and time-dependent density functional theory (TD-DFT) calculations use the same method. All calculations used 6-311++G(d,p) basis set (BS) with the implicit polarizable continuum model (PCM).[16] That is, AG-1478 geometries at the ground electronic state ($S_0$) and first singlet excited-state ($S_1$) in Figure 1 were optimized, using the LR-PCM/TD-DFT methods in dimethyl sulfoxide (DMSO) solvent, followed by TD-DFT calculations using the same DFT functionals (Vxc) in DMSO. The 22 DFT functionals (21 DFT functional plus HF) are available from the Gaussian 16 computational chemistry software packages,[17] and Table S1 summarize the calculated results using different DFT Vxc functionals.

For the investigation of absorption and emission processes, two approaches are usually employed: the conventional linear-response (LR) approaches[18] and the state-specific (SS) ones.[19] In the LR method, the absorption and emission energies including a PCM correction are determined by the electron density variation associated with the transition. All calculations were performed using Gaussian 16 computational chemistry packages[17] at the supercomputing facilities from National Computational Infrastructure (NCI) and the OzSTAR Supercomputer at Swinburne University. The AG-1478 chemical structure in Figure 1b is the global minimum structure obtained based on the B3LYP/6-311++G(d,p) model.[3] Analysis of the linear response of the ground-state density was calculated either with the HF or with DFT to an external time-dependent perturbation leads to the TD-HF or TD-DFT schemes, respectively,[11] where the wavefunction Hartree-Fock (HF) calculations are employed as a reference.

## 3. RESULTS AND DISCUSSION

### 3.1 Performance of the DFT-VXC functionals in TD-DFT study

Although several general conclusions are available, there is no universal solution for choosing the "best" $V_{XC}$ for the TD-DFT calculations of all systems of interest.[12] As a result, it becomes a common practice of benchmarking studies to assess the $V_{XC}$ for calculating particular properties, (e.g., absorption and/or emission spectra) of a particular class of compounds (e.g., quinazoline scaffold TKIs in our case). A comprehensive review and extensive assessment of about 200 DFT-$V_{XC}$ functionals by Mardirossian and Head-Gordon[13] divided the DFT-$V_{XC}$ functionals approximately into three major corrections of wave function theory, exchange-correlation, and dispersion corrections. As many as eight energetic properties were studied such as non-covalent "easy" energy (NCED, NCEC, NCD), isomerization energy (IE, ID), thermochemistry energy (TCE), and barrier height (BH). The top 20 DFT-$V_{XC}$ functional performers included SPW92, PBE, TPSS, B3LYP; PBE-D3(BJ), revPBE-D3(BJ), BLYP-D3(BJ), B97-D3(BJ); TPSS-D3(BJ), SCAN-D3(BJ), M06-L, B97M-RV; PBE0-D3(BJ), B3LYP-D3(BJ), ωB97X-D, ωB97X-V; TPSSh-D3(BJ), M06-2x, MN15 and ωB97M-V.[13] In this assessment, Perdrews' metaphorical Jacob's Ladder with five rungs of accuracy order of local spin-density approximation (LSDA), generalized gradient approximation (GGA), meta-GGA, hybrid GGA/meta-GGA, and double hybrid functionals was discussed. However, all top DFT-$V_{XC}$ functionals recommended by Mardirossian and Head-Gordon[13] were for ground electronic states and their energetic properties rather than spectroscopic properties in excited states. In addition, the training database consisted of relatively small molecules.

Figure 1 gives the chemical structure, nomenclature (Figure 1a) and the optimized geometries of the ground (Figure 1b) and first excited states (Figure 1c) of AG-1478 using B3LYP/6-311++G(d,p). Table S1 summarizes the maximum absorption ($\lambda_{ab}$) and emission ($\lambda_{em}$) values of AG-1478 calculated using 21 DFT-$V_{XC}$ functionals employed in the present TD-DFT study, together with results from the time-dependent Hartree-Fock (TD-HF) scheme. All calculations are performed using DFT-$V_{XC}$/6-311++G (d, p) in the same manner, which is detailed in the computational details section. The TD-HF results were used as a reference for the present benchmarking study.

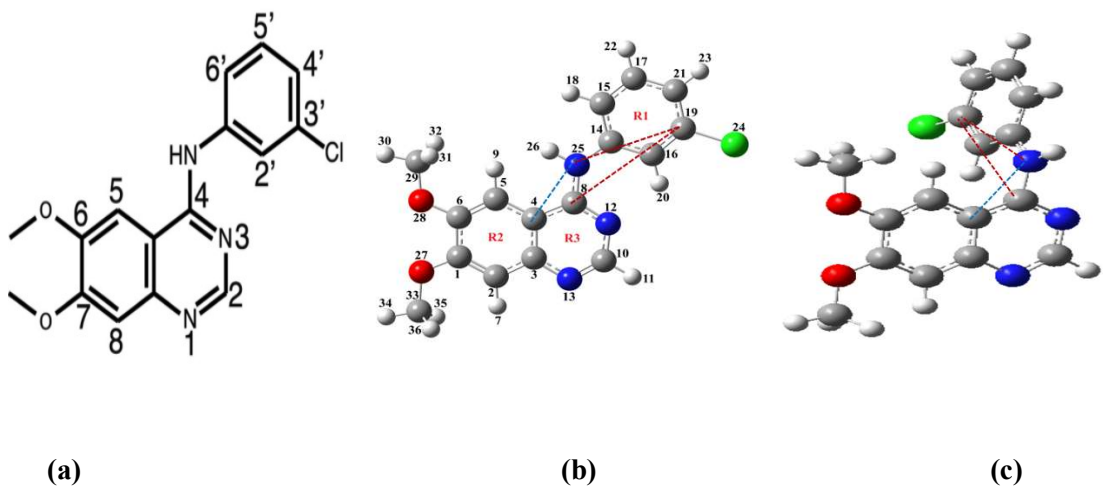

(a)  (b)  (c)

**Figure 1.** The chemical structure of AG-1478 (left with IUPAC nomenclature) (a), optimized ground electronic state structure (b), and optimized first excited structure of AG-1478. Note the dihedral angle, ∠$C_{(10)}$-$C_{(4)}$-NH-$C_{(3')}$ (IUPAC) marked in the three-dimensional (3D) structures (b) and (c) indicates the planar and twisted orientational of the chromone rings. The numbers in (b)

are the Gaussian labeling of the atoms with R1, R2, and R3 are perimeters for the corresponding rings of the quinazoline scaffold.

At present, TD-DFT represents one of the most prominent approaches for optical reporting studies, especially for excited states of medium-sized and large molecular systems such as AG-1478.[11] Figure 2 compares the individual performance (i.e. the discrepancies in eV) between the calculated and the measured[4] maximum optical wavelengths ($\lambda_{max}$) of AG-1478 in DMSO using 21 different DFT methods as well as the HF. The optical properties including maximum absorption (blue, $\Delta\lambda_{ab} = \lambda_{ab}^{cal} - \lambda_{ab}^{exp}$), emission (orange, $\Delta\lambda_{em} = \lambda_{em}^{cal} - \lambda_{em}^{exp}$) and Stokes shifts (grey, $\Delta\lambda_{ss} = \lambda_{ab}^{cal} - \lambda_{em}^{exp}$), and overall error (yellow, $\lambda_{overall} = |\Delta\lambda_{ab}| + |\Delta\lambda_{em}| + |\Delta\lambda_{ss}|$) of AG-1478 are presented in Table S1. Note the experimental maximum absorption used is the $\lambda_{max}$ (i.e., $\lambda_2$) not $\lambda_1$. As seen in Figure 2, the performance of a DFT $V_{XC}$ functional depends on the accuracy of the calculated absorption and emission, as well as the Stokes shift. In the present study In the present study for practical reasons,[4] the Stoke shifts are obtained using the maximum wavelengths ($\lambda_{max}$) of the calculated absorption and emission spectra.[20] Note that at room temperature, AG-1478 can exist in the form of multiple conformers.[5]

We also calculated an overall error $\Delta\lambda$ which is the sum of absolute values of all three properties (the yellow bar in Figure 2). It is very clear in Figure 2 that the accuracy of the DFT-$V_{XC}$ functionals can be measured using the overall error $\Delta\lambda$ because just two inaccurate wavelengths $\lambda_{em}$ and $\lambda_{ab}$ may produce an "accurate" Stokes shift due to the error cancellation. For example, In Figure 2 shows that the HF method and LC-B3LYP produce larger errors in both $\lambda_{em}$ (blue) and $\lambda_{ab}$ (orange) but an error in Stokes shift (grey) is small due to the cancellation.

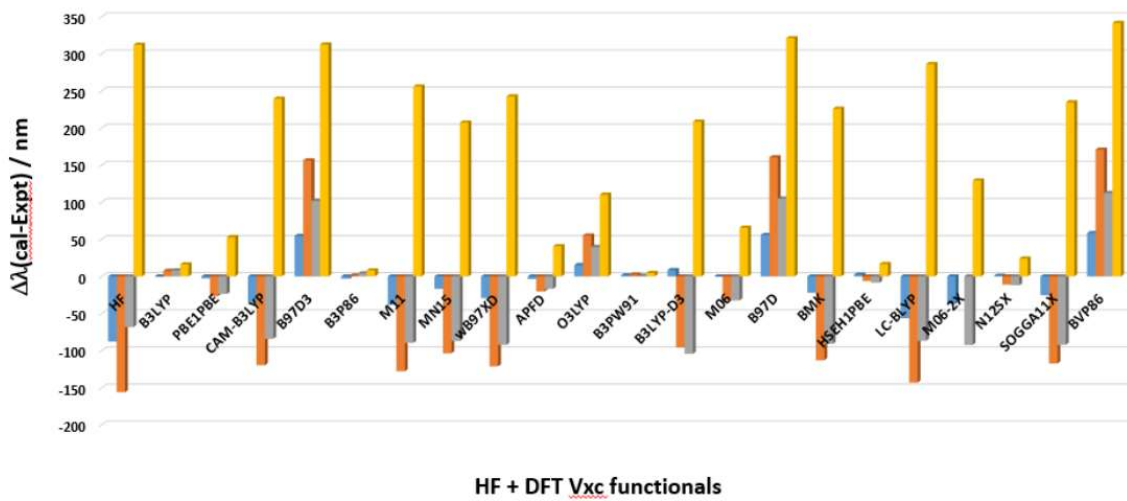

**Figure 2.** Performance of DFT-$V_{XC}$ functionals with respect to producing the accurate maximum absorption wavelength (blue, $\Delta\lambda_{ab} = \lambda_{ab}^{cal} - \lambda_{ab}^{exp}$ (332nm)), emission wavelength (orange, $\Delta\lambda_{em} = \lambda_{em}^{cal} - \lambda_{em}^{exp}$ (475.5nm)), Stokes shift (grey, $\Delta\lambda_{ss} = \lambda_{ab}^{cal} - \lambda_{em}^{exp}$), and overall error (yellow, $\Delta\lambda_{overall} = |\Delta\lambda_{ab}| + |\Delta\lambda_{em}| + |\Delta\lambda_{ss}|$) of AG-1478 in DMSO (nm). Details are presented in Table S1 of the supplementary materials.

To compare the DFT methods, the results obtained using the TD-HF method demonstrate large errors for both absorption and emission wavelengths calculated, as shown in Figure 2 and Table S1. Figure 2 shows that DFT functionals such as BVP86 and B3PW91 are two extremes of the performance of poor and excellent, respectively. Most of the DFT $V_{XC}$ functionals result in errors in $\Delta\lambda_{ab}$ and $\Delta\lambda_{em}$ in the same direction, i.e., overestimates or underestimates both absorption and emission energies (wavelengths). One needs to note that three DFT-$V_{XC}$ functionals, B3LYP, B3P86, and B3PW91 are among the top performers in terms of calculation of the optical spectra of AG-1478. Interestingly, Becke's three-parameter (B3) with correlation Vxc DFT functionals, B3Vc, also produce accurate excited state energies (wavelengths) using TD-DFT. For example, the errors in absorption wavelength $\Delta\lambda_{ab}$ using B3LYP, B3PW91 and B3P86 are -0.63 nm, 1.39 nm, and -3.01 nm, respectively, whereas the errors in emission wavelength $\Delta\lambda_{em}$ using the same B3Vc functionals are 7.43 nm, 2.28 nm, and 1.1 nm, accordingly. These B3Vc functionals are reliable tools for the study of the optical properties of the TKIs and therefore the design of new agents with larger Stokes shift for medical image applications.

The success of the Becke three-parameter DFT functionals (B3LYP, B3PW91, and B3P86) has been recognized by earlier studies for small molecules. They are, however, not exactly as the notation stands for, if these functionals are employed in the Gaussian 16 computational chemistry packages (Fisch, et al., 2016) for the TD-DFT calculations like the present study. In 1993, Becke proposed three-parameter hybrid functional B3WP91,[21]

$$E_{XC}^{B3PW91} = E_{XC}^{LDA} + a(E_X^{exact} - E_X^{LDA}) + b\,\Delta E_X^{B88} + E_C^{VWN} + c\,\Delta E_C^{PW91} \quad (1)$$

where $a$=0.20; $b$=0.72 and $c$=0.81. The three parameters $a$, $b$, and $c$ were fit to the atomization-energy data, which were determined by Becke via fitting to the G1 molecule set.[22] As a result, the B3PW91 correlation functional has contributions from the VWN functional[23] and c*PW91 functional[24] where **c** is fit by Becke. In 1994, Frisch and co-workers[25] reworked the B3PW91 in Eq (1) using the Lee, Yang, and Parr (LYP) functional[26] for correlation density-functional approximation (DFA) instead of the PW91 functional. The reworked functional employs the same three parameters as in Eq. (1), which is known as B3LYP. Similarly, the B3P86 functional was reworked. The Gaussian programs[17] employ the same $a$, $b$, and $c$ values for the LYP, P86, and PW91, and introduced the VWN functional III for (VMN III)[23] for the local contribution. As a result, such reworked B3Vc functionals are excellent performers for AG-1478 optical spectra.

Interestingly, the B3LYP functional performs very well for both absorption and emission wavelengths of AG-1478 in DMSO. Any modifications to the "original" B3LYP functional, such as CAM-B3LYP and B3LYP-D3, largely reduce the accuracy in the emission wavelength (excited states) calculations. For example, the CAM-B3LYP functional reduces the accuracy in the absorption wavelength ($\Delta\lambda_{ab}$) calculation from -0.63 to -35.7 nm while the B3LYP-D3 slightly improves the accuracy of absorption wavelength ($\Delta\lambda_{ab}$) calculation from -0.63 to 8.45 nm. As a result, the B3LYP DFT functional is the preferred one over its derivatives such as CAM-B3LYP and B3LYP-D3 for the optical properties calculations, though Grimme's dispersion correction DFT-D3 approach[27] is satisfactorily accurate at practically the same computational cost as pure DFT for small systems.[28]

The performance of the DFT-$V_{XC}$ functionals, along with HF, can be grouped into poor (overall error $\Delta\lambda > 150$ nm), reasonable (50 nm $< \Delta\lambda <$ 150 nm), and excellent ($\Delta\lambda <$ 50 nm). The electron correlation energy seems to be important for optical properties and all DFT-$V_{XC}$ functionals include the electron correlation energy in varying degrees. When paired with the same 6-311++G (d, p) base set, the TD-HF shows a very large error of nearly $\Delta\lambda \sim 300$ nm. The DFT-$V_{xc}$ functionals with poor performance include CAM-B3LYP, B97D3, M11, NM11, ωB97XD, B3LYP-D3, B97D, BMK, LC-B3LYP, SOGGA11X, and BVP86 with an overall error of $\Delta\lambda >$ 150 nm. For example, the LC-BLYP is not suitable for studying the optical properties of the TKI because it gives a very large overall error of over, $\Delta\lambda >$ 286 nm. A group of DFT-$V_{XC}$ functionals yielding moderate accuracy includes PBE1PBE, O3LYP, and M06. Many of them produce small absorption ($\Delta\lambda_{ab}$) errors but poor emission ($\Delta\lambda_{em}$) ones. As shown in Figure 2, almost all orange columns (error in an excited state prediction) are larger than the corresponding blue ones (errors in absorption prediction).

As shown in Figure 2, the top DFT-$V_{XC}$ functional performers which accurately produce the optical properties of AG-1478 are B3LYP, PBE1PBE, B3P86, APFD, B3WP91, HSEH1PBE, and N12SX. They can produce the optical properties within a range of 0.1-0.5 eV, which is comparable with the accuracy of high-level correlated approaches such as EOM-CCSD or CASPT2,[11] with the (EOM-CCSD and CASPT2) having significant computational costs.

The top-performing DFT-$V_{XC}$ functionals demonstrate small errors of $\Delta\lambda<50$ nm and they are shown in Figure 3. Our further discussion will concentrate on the best performing DFT-$V_{XC}$ functionals together with HF calculations as a reference.

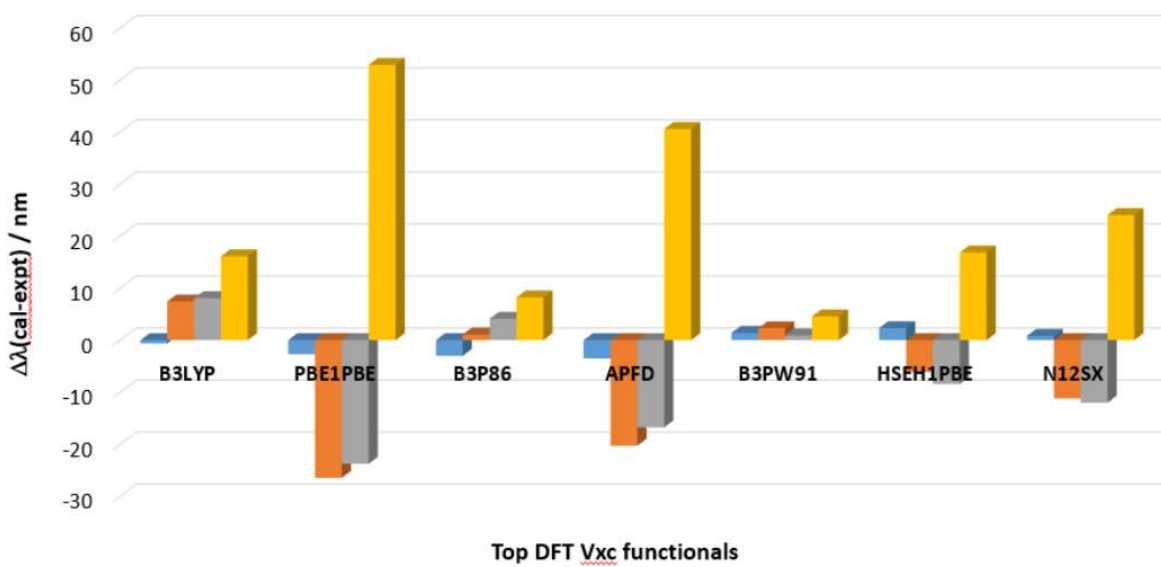

**Figure 3.** Top seven DFT-$V_{XC}$ functional performers with respect to accurately producing maximum absorption (blue, $\Delta\lambda_{ab} = \lambda_{ab}^{cal} - \lambda_{ab}^{exp}$), emission (orange, $\Delta\lambda_{em}=\lambda_{em}^{cal} - \lambda_{em}^{exp}$), Stokes shift (grey, $\Delta\lambda_{ss}=\lambda_{ab}^{cal} - \lambda_{em}^{exp}$ ), and overall error (yellow, $\lambda_{overall} = |\Delta\lambda_{ab}| + |\Delta\lambda_{em}| + |\Delta\lambda_{ss}|$) of AG-1478 in DMSO.

Performance of the DFT-$V_{XC}$ functionals in term of overall error ($\lambda_{overall} = |\Delta\lambda_{ab}| + |\Delta\lambda_{em}| + |\Delta\lambda_{ss}|$) for AG-1478 in DMSO (the smaller the better) is given as B3PW91 (4.56 nm) > B3P86 (8.22 nm) > B3LYP (16.12 nm) > HSEH1PBE (16.84 nm) >N12SX (24.04 nm) > AFPD (40.58 nm) > PBE1PBE (52.9 nm). The results are in agreement with the recent benchmarking study of DFT-

$V_{XC}$ functionals using TD-DFT calculations for the UV-Vis spectral simulation[14] who obtained the following order in terms of accuracy: B3LYP > APDF > M06 > PW6B9D3 > PBE0 > BP86 > PBE > M06-2X > CAM-B3LYP > ωB97XD > LC-ωPBE.

**3.2 Ground state and first excited state of AG-1478 insolvent**

For studying the ground state (GS, $S_0$) and excited state (ES) properties it is very important to choose the proper calculation method. Within the framework of the TD-DFT, for calculations of the absorption wavelengths, one needs to estimate the excited state (ES) energies (and related properties) using the optimal ground-state (GS) geometry without exploring the ES potential energy surface. This approach is usually a good approximation for absorption (eg UV-vis) calculations.[3,14,29-31] The GS properties include geometry, dipole moment, and electronic transitions (absorption).[32] If not using the high-level CASSCF/CASPT2 approaches, proper selection of the most suitable DFT-$V_{XC}$ functional(s) depends on the molecular scaffold.[33-35] In the case of accurate but time-consuming CASSCF/CASPT2 methods, some studies are limited by the size of the molecular system. For example, for studying the electronic structure properties of the GS and ES of large drugs such as lapatinib, Vaya et al.[36] restricted their study to the core scaffold containing only the furan and quinazoline chromophores because of the high CPU cost of the CASSCF/CASPT2 methods. However, under this approach, the important effects of side groups are not considered.

Optimizations of the GS and the first ES using the DFT-$V_{XC}$ functionals converges to the planar and twisted structures of AG-1478 presented in Figure 1b and 1c, respectively. Tables S2 and S3 compare the deviations of the calculated geometric perimeters of the phenyl ($R_1$) and the quinazoline rings $R_2$ and $R_3$, together with the C-Cl bond length of AG-1478 in DMSO using all DFT-$V_{XC}$ functionals for the GS and the first ES, respectively. The perimeters of the all-carbon rings ($R_1$ and $R_2$) share some similarities with close perimeters of 8.459 Å and 8.480 Å, in agreement with the earlier study of AG-1478.[3] The perimeter of the pyrimidine ring ($R_3$) of the quinazoline moiety is smaller in AG-1478 than the all-carbon ring perimeters.

ES's often exhibit different geometries from the GS's, which can be attributed to the insurgence of the solvatochromic behavior to the presence of a twisted intermolecular charge transfer (TICT) state that may be accessible upon excitation.[37] In addition, the optimized geometric parameters such as ring perimeters in the first excited state of AG-1478 are not the same as their corresponding perimeters in the ground state using the same method. For example, the perimeters of $R_1$, $R_2$, and $R_3$ are 8.372 Å, 8.442 Å, 8.217 Å for the GS, respectively, which become 8.414 Å, 8.452 Å, and 8.267 Å for the ES using B3LYP/6-311++G(d,p). The same parameters are 8.351 Å, 8.418 Å, 8.186 Å in the GS, which becomes 8.390 Å, 8.426 Å, and 8.236 Å for the ES when PBE1PBE/6-311++G (d, p) is employed. The C-Cl bond length is 1.769 Å in the GS but shorter (1.747 Å) in the first ES using the same B3LYP/6-311++G (d, p) method. This is in agreement with the B3LYP/6-311G(d,p) and the experimental study of flufenpyr and amipizone.[38]

Table 1 compares the data obtained from the optimized GS ($S_0$) and the first ES ($S_1$) using the top-performing DFT-$V_{XC}$ functionals. The results from the DFT and HF calculations indicate that the GS of AG-1478 in DMSO is a planar configuration, in agreement with earlier studies.[3] The dihedral angle, $\delta = \angle C_{(10)}$-$C_{(4)}$-NH-$C_{(3')}$ (refer to Figure 1) which determines the relative orientation of the quinazoline and chloro-phenyl planes is nearly planar and equal to 180°, with the dipole moment being about 8.0 D, except for HF (7.31 D). The calculated maximum absorption wavelength $\lambda_{ab}$ is close to the experimental absorption wavelength of 332 nm except for the HF of (244.31 nm). Note that the calculations are only based on the global minimum structure of AG-1478, while it is known that other conformers can be populated under room temperature.[3,5]

**Table 1.** Ground state and the first excited state properties according to the top seven DFT-$V_{XC}$ performers in DMSO*.

| Method | Ground State ($S_0$) | | | | | |
|---|---|---|---|---|---|---|
| | μ / D | $R_1$ /Å | $R_2$ /Å | $R_3$ /Å | C-Cl /Å | δ=∠$C_{(10)}$-$C_{(4)}$-NH-$C_{(3')}$*/° |
| HF | 7.31 | 8.124 | 8.383 | 8.316 | 1.752 | -179.97 |
| B3LYP | 8.08 | 8.372 | 8.442 | 8.217 | 1.769 | -180.0 |
| B3P86 | 8.06 | 8.351 | 8.418 | 8.191 | 1.752 | -179.98 |
| B3PW91 | 8.06 | 8.360 | 8.428 | 8.199 | 1.754 | -179.98 |
| BPE1BPE | 8.00 | 8.351 | 8.418 | 8.186 | 1.748 | -180.0 |
| APFD | 8.05 | 8.358 | 8.423 | 8.190 | 1.749 | -179.99 |
| HSEH1PBE | 8.03 | 8.352 | 8.418 | 8.187 | 1.749 | -179.83 |
| N12SX | 7.97 | 8.328 | 8.395 | 8.161 | 1.747 | -179.82 |
| Method | 1st Excited State ($S_1$) | | | | | |
| | μ / D | $R_1$ /Å | $R_2$ /Å | $R_3$ /Å | C-Cl /Å | δ=∠$C_{(10)}$-$C_{(4)}$-NH-$C_{(3')}$*/° |
| HF | 7.34 | 8.325 | 8.478 | 8.196 | 1.752 | -153.50 |
| B3LYP | 6.90 | 8.414 | 8.452 | 8.267 | 1.747 | -98.66 |
| B3P86 | 6.81 | 8.391 | 8.429 | 8.239 | 1.732 | -95.73 |
| B3PW91 | 6.84 | 8.400 | 8.437 | 8.247 | 1.734 | -96.37 |
| BPE1BPE | 6.81 | 8.390 | 8.426 | 8.236 | 1.729 | -97.09 |
| APFD | 6.00 | 8.398 | 8.435 | 8.236 | 1.730 | -79.01 |
| HSEH1PBE | 6.02 | 8.394 | 8.430 | 8.235 | 1.729 | -96.58 |
| N12SX | 6.78 | 8.371 | 8.408 | 8.208 | 1.727 | -96.89 |

*using 6-311++G (d, p) basis set. The HF calculations are as a reference.

Table 1 shows data for the first singlet ES (S$_1$) of AG-1478. The optimized structure of the ES is a twisted configuration (see Figure 1c). Dihedral angle δ=∠C$_{(10)}$-C$_{(4)}$-NH-C$_{(3')}$ in the ES is approximately -96°. The same angle from the HF calculations is larger (at δ = -153.5°) but the APFD yields a smaller value ( δ = -79.01°). The twisted configuration of the ES of AG-1478 leads to the decrease of the dipole moment, from 7-8 D in the GS to 6-7 D in the first ES. Figure 4 illustrates alignments of all optimized ES structures of AG-1478 using the top-performing DFT-V$_{XC}$ functionals.

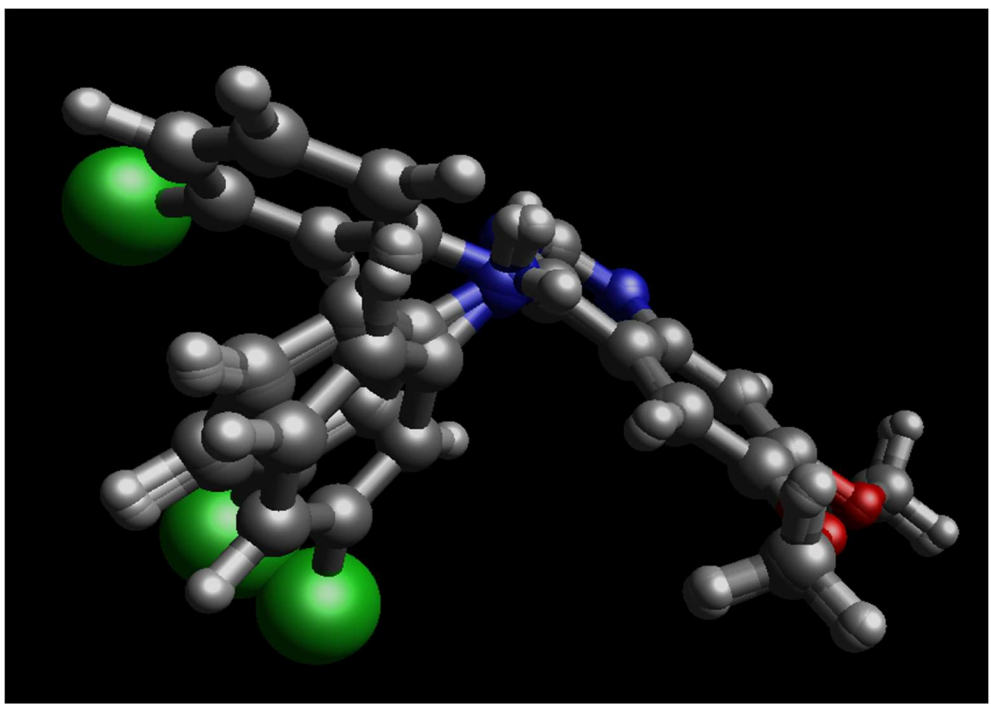

**Figure 4.** Alignment of all optimized 1$^{st}$ ES structures of AG-1478 using B3LYP, B3P86, B3PW91, BPE1BPE, APFD, HSEH1PBE, and N12SX DFT-V$_{XC}$ functions in DMSO. The 6-311++G(d, p) basis set was used for all calculations. The HF method gave δ=-153.50° while the APFD yielded the smallest dihedral angle δ=-79.01°.

The states relevant for the absorption and emission are the two lowest singlet states based on the ground state (S0) and first excited state (S1) geometries of AG-1478 in Figures 1b and 1c. The major configurations of absorption and emission are dominated by transitions between the highest occupied molecular orbital (HOMO, MO82a) and the lowest unoccupied molecular orbital (LUMO, 83a), i.e., the HOMO → LUMO transitions. Figure 5 reports the absorption wavelength $\lambda_{ab}$ and emission wavelength $\lambda_{em}$ of AG-1478, respectively, between the orbitals, HOMO (82a) and (LUMO,83a) of AG-1478 in DMSO. The absorption transition of HOMO-LUMO represents a local excitation (LE) on the chromone rings, with minor charge transfer (CT) on the Cl atom. However, as shown in Figure 5b, the emission can be characterized as a twisted intramolecular charge transfer (TICT) in which charge transfers from the phenyl to the quinazoline chromone, in agreement with a study of fluorescence spectra of isoflavones by Beyhan et al.[7] and a study of excited states of 4-(N, N-Dimethylamino)-4′- nitrostilbene.[39]

(a) Ground state (S$_0$)

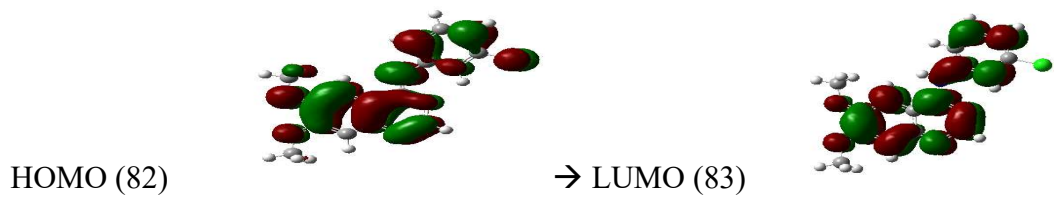

HOMO (82)     → LUMO (83)

(b) First excited state (S$_1$)

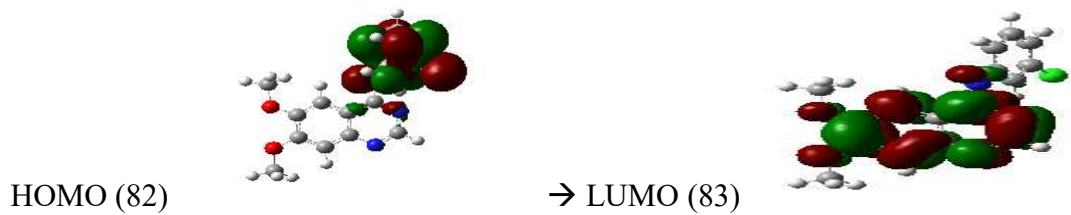

HOMO (82)     → LUMO (83)

**Figure 5.** HOMO→LUMO transitions calculated using the TDDFT calculations with B3LYP/6-311++G(d, p) for (a) absorption ($\lambda_{ab}$ = 331.37 nm) as a local excitation (LE) and (b) emission ($\lambda_{em}$= 482.93 nm) as twisted intramolecular charge transfer (TICT) of the AG-1478 in DMSO solvent.

The lowest excited state (ES) is often the most photobiological relevant state for biomolecules. Low excited states are less likely with charge-transfer problems and without significant Rydberg character so that range-separated functionals are not generally necessary.[6] To understand the experimentally observed maximum emission wavelength at 475.5 nm (2.607 eV)[3] of AG-1478 in more detail, we consider the first singlet excited state. The corresponding adiabatic excitation energies and excited state dipole moments were calculated using the top-performing DFT-$V_{XC}$ functionals and are given in Table 2, together with a dihedral angle δ between the rings. Relaxation of the lowest excited singlet states is always accompanied by a significant rotation toward planarity around the chromone-phenyl bond relative to the ground state structure. The largest change thereby occurs in the TICT state, with the dihedral angle of only δ being only ≈-96° at its equilibrium geometries. According to the measured spectra of AG-1478,[4] the absorptions ($\lambda_1$ and $\lambda_2$) occur at 346 nm and 332 nm in DMSO, respectively, and the emission occurs at 475.5 nm.

**Table 2.** Ground state and the first excited state properties were calculated using the top seven DFT-$V_{XC}$ performers in DMSO*.

| DFT-Vxc | λab / nm | Δλab / nm | λex / nm | λem / nm | Δλem / nm | Δλss#= λem- λab / nm | Δ(Δλss) / nm | Sum (\|Δλ\|) / nm |
|---|---|---|---|---|---|---|---|---|
| HF | 244.31 | -87.69 | 236.24 | 319.47 | 156.03 | 75.16 | -68.34 | 312.06 |
| B3LYP | 331.37 | -0.63 | 319.56 | 482.93 | 7.43 | 151.56 | 8.06 | 16.12 |
| PBE1PBE | 329.30 | -2.70 | 315.67 | 449.05 | -26.45 | 119.75 | -23.75 | 52.90 |
| B3P86 | 328.99 | -3.01 | 316.64 | 476.6 | 1.10 | 147.61 | 4.11 | 8.22 |
| APFD | 328.47 | -3.53 | 307.91 | 455.21 | -20.29 | 126.74 | -16.76 | 40.58 |
| B3PW91 | 333.39 | 1.39 | 310.82 | 477.78 | 2.28 | 144.39 | 0.89 | 4.56 |
| HSEH1PBE | 334.32 | 2.32 | 311.24 | 469.4 | -6.10 | 135.08 | -8.42 | 16.84 |
| N12SX | 332.81 | 0.81 | 310.63 | 464.29 | -11.21 | 131.48 | -12.02 | 24.04 |

| Expt | 332 | - | - | 475.5 | - | 143.5 | - | |

*using 6-311++G (d, p) basis set. The HF calculations serve as a reference.

#The Stokes shift was calculated as the difference between the maximum emission ($\lambda_{em(max)}$) and maximum absorption ($\lambda_{ab(max)}$). This experimental value is 475.5-332=143.5 nm. Note that the experiment Stokes shift is calculated using the first absorption ($\lambda_1$) and maximum emission ($\lambda_{em}$).[4]

Table 2 shows the maximum absorption and emission wavelengths, the Stokes shifts, and the overall error of the top DFT functionals. Some DFT functionals produce accurate absorption wavelengths, some are good for the emission wavelengths, some are accurate for Stokes shifts and some achieve overall small errors. For example, the top three DFT functional for absorption wavelengths are B3LYP (-0.63 nm), N12SX (0.81 nm), and B3PW91 (1.39 nm). The top three DFT functional for prediction of the emission wavelengths are not necessarily the same as absorption but B3P86 (1.10 nm), B3PW91 (2.28 nm), and HSEH1PBE (-6.10 nm). As for the Stokes shifts, the top three performers are B3PW91 (0.89 nm), B3P86 (4.11 nm), and B3LYP (8.06 nm). Finally, the top three DFT functionals in terms of accuracy (the smaller the better) are B3PW91 (4.56 nm), B3P86 (8.22 nm), and B3LYP (16.12 nm). Thus, B3PW91 functional is the top overall performer for prediction of the optical properties including absorption, emission, and Stokes shifts. The B3LYP is excellent for absorption calculations. B3P86 is the best for emission calculations.

4. CONCLUSION

Performance of twenty-one DFT functionals and the Hartree-Fock (HF) theory was benchmarked with respect to the measured optical properties of TKI AG-1478 in dimethyl sulfoxide (DMSO) using the TD-DFT method. The Becke's three-parameter functionals,

B3PW91, B3LYP, and B3P86, combined with the 6-311++G(d, p) basis set demonstrated the top overall performance in reproducing the experimental optical properties. Other DFT functionals such as BPE1BPE, APFD, HSEH1PBE, and N12SX are also among the excellent performers. Although B3LYP functional is a top performer, any corrections to it, such as CAM-B3LYP, LC-B3LYP, and B3LYP-D3, resulted in larger errors in the optical properties of AG-1478 in DMSO. More specifically, B3PW91 is a recommended DFT functional for studying the optical properties of the TKI class of compounds. B3LYP is an excellent method for absorption calculations while B3P86 is the best for emission calculations. These B3Vc functionals are reliable tools for studying the optical properties of the TKIs and therefore for the design of new agents with larger Stokes shift for medical image applications.

Significant research has been dedicated to the improvement and development of new $V_{XC}$ functionals to eliminate the known failures of TD-DFT.[11] However, a big remaining question is that whether there exists the approximate $V_{XC}$ functional that can describe both the ground state and excited states equally well. Since different excited states can possess very different electronic structures, it seems unlikely that all of them can all be captured by a single approximate $V_{XC}$ functional for all classes of molecules. The present benchmarking shows that some DFT functionals may be accurate for the ground electronic state but not for the lowest excited state or vice versa.[11] Thus, it is important to benchmark the DFT functionals for a particular class of molecules and desired properties before production calculations.

AUTHOR INFORMATION

**Corresponding Author**

*E-mail: fwang@swin.edu.au

**Author Contributions**

The manuscript was written through contributions of all authors. All authors have given approval to the final version of the manuscript.

ACKNOWLEDGMENT

SA acknowledges Swinburne University of Technology Tuition Fee Scholarship (TFS). FW acknowledges part of the funding from Excellerate Australia for "Spectroscopic and theoretical study of a potent anticancer drug." The authors acknowledge supercomputer support from the National Computational Infrastructure (NCI) at Australian National University (ANU) and from Swinburne University of Technology Supercomputing Facilities (OzSTAR).